\documentclass[twocolumn,prb,showpacs,showkeys,preprintnumbers,amsmath,amssymb]{revtex4}
\usepackage{graphicx}

\begin{document}

\title{Variation of metal-insulator transition and formation of bipolarons by the Cd-doping in the thiospinel
 system of Cu$_{1-x}$Cd$_{x}$Ir$_{2}$S$_{4}$}

\author{Guanghan \surname{Cao}}

\thanks{Author to whom correspondence should be addressed}

\email{ghcao@zju.edu.cn}

\affiliation{Department of Physics, Zhejiang University, Hangzhou
310027, P. R. China}
\affiliation{ National Institute for Materials
Science, Sengen
1-2-1, Tsukuba, Ibaraki 305-0047, Japan}%

\author{Hideaki Kitazawa, Takehiko Matsumoto}

\affiliation{ National Institute for Materials Science, Sengen
1-2-1, Tsukuba, Ibaraki 305-0047, Japan}%

\author{Chunmu Feng}
\affiliation{Test {\&} Analysis Center, Zhejiang University,
Hangzhou 310027, P. R. China}

\begin{abstract}
The pseudo-binary thiospinel system
Cu$_{1-x}$Cd$_{x}$Ir$_{2}$S$_{4}$ was investigated by the x-ray
diffraction, electrical resistivity, magnetic susceptibility and
specific heat measurements. It was shown that the system exhibits
a miscibility-gap behavior for the Cd substitution, however,
nearly monophasic samples was obtained by quenching at 1373 K,
except for $0.4< x \leqslant 0.6$. With increasing the Cd
concentration, the room-temperature electrical conductivity and
Pauli susceptibility decrease monotonically, consistent with the
hole-filling picture. The first-order metal-insulator transition
at about 230 K in the parent compound CuIr$_{2}$S$_{4}$ is changed
into a second-order transition around 185 K when $x\sim$ 0.25,
whereafter the second-order transition disappears at $x\sim 0.8$.
No superconductivity was observed down to 1.8 K. The end-member
compound CdIr$_{2}$S$_{4}$ is shown as an insulator with a
band-gap of 0.3 eV.  Analysis for the data of magnetic
susceptibility and electrical resistivity suggests the formation
of bipolarons below 185 K for $0.25 <x<$ 0.8, which accounts for
the absence of superconductivity in terms of the transition from
the BCS Cooper pairs to small bipolarons.
\end{abstract}



\maketitle

\section{INTRODUCTION}

Thiospinel CuIr$_{2}$S$_{4}$ has recently attracted much interest
for its novel metal-insulator (MI)
transition.~\cite{Nagata94,Hagino,
Furubayashi94,Oomi,Suzuki,Cao01,Radaelli,Oda,Kumagai,Matsuno} The
MI transition takes place at about 224 K on cooling, accompanied
by a sudden increase of the electrical resistivity, a quenching of
the Pauli magnetic susceptibility, and a structural phase
transition towards lower symmetry.~\cite{Nagata94,Hagino,
Furubayashi94} It was found that the MI transition temperature
\emph{increases} with applying
pressures.~\cite{Furubayashi94,Oomi} Recently, we found that the
Zn substitution for Cu resulted in appearance of superconductivity
by the suppression of the MI transition.~\cite{Suzuki, Cao01}

The mechanism of the MI transition has remained an open question.
We have elucidated it in terms of the interatomic Coulomb
interactions and the dimerizations of Ir$^{4 + }$ based on the Zn
substitution result.~\cite{Cao01} Very recently, the
low-temperature insulating phase was revealed to be charge ordered
and spin-dimerized,~\cite{Radaelli} consistent with our proposal.
However, further studies are needed to clarify the driving force
of the unusual spin-dimerization in a three dimensional compound.

In this paper, we studied the Cu$_{1-x}$Cd$_{x}$Ir$_{2}$S$_{4}$
system based on the following considerations. First, the cadmium
substitution at the A-site of the normal spinel AB$_{2}$X$_{4}$
hardly changes the IrS$_{6}$-octahedron framework, which
determines the electronic structure of the valence
bands.~\cite{Oda} Second, the valence state of Cu has been
confirmed to be 1+,~\cite{Kumagai,Matsuno} and the energy level of
Cu 3$d^{10}$ or Cd 4$d^{10}$ is far below the Fermi energy
$E_{F}$. Therefore, the Cd-substitution is expected to alter the
carrier concentration like the Zn-substitution does in Cu$_{1 -
x}$Zn$_{x}$Ir$_{2}$S$_{4}$. Third, the ionic radius of Cd$^{2+}$
is obviously larger than that of Cu$^{1+}$ or
Zn$^{2+}$.~\cite{Shannon} So, the Cd substitution will bring in a
negative chemical pressure for the lattice, which is expected to
affect both of the charge ordering and the spin dimerization.
Finally, one would be interested in whether or not
superconductivity could also appear in the Cd-substitution system.

\section{EXPERIMENTS}

Samples of Cu$_{1-x}$Cd$_{x}$Ir$_{2}$S$_{4}$ ($x$=0, 0.025, 0.05,
0.1, 0.15, 0.2, 0.25, 0.3, 0.4, 0.5, 0.6, 0.7, 0.8, 0.9 and 1.0)
were prepared by the chemical reaction in a sealed quartz ampoule.
The starting material was Cu, CdS, Ir and S powders with the
purity higher than 99.9{\%}. The sealed ampoule was first heated
slowly to 1023 K, and then fired at $\sim $1373 K for 96 hours. In
order to increase the Cd solubility, samples were quenched. The
as-prepared powder was pressed into pellets with the pressure of
2000 kg/cm$^{2}$, and the pellets were sealed and sintered at the
same temperature for 48 hours and then quenched.

Powder x-ray diffraction (XRD) was carried out at room temperature
with CuK$\alpha $ radiation by employing a RIGAKU X-ray
Diffractometer. The crystal structure was refined by the RIETAN
Rietveld analysis program.~\cite{Izumi} The electrical resistivity
($\rho )$ was measured by the standard dc four-probe method down
to 2 K. The magnetic susceptibility ($\chi$) was measured by using
a Quantum Design SQUID magnetometer, whose precision achieves
$\sim 10^{-9}$ emu. The measurement was carried out using about
100 mg samples in both cooling and heating processes between 1.8 K
and 300 K under applied field of 1000 Oe. The background was
measured in advance and then subtracted, so that the net $\chi$
value could be accurately obtained. Finally, the heat capacity was
measured on a Quantum Design PPMS system using an automated
relaxation technique.

\section{RESULTS AND DISCUSSION}

\begin{figure}
\includegraphics[width=7cm]{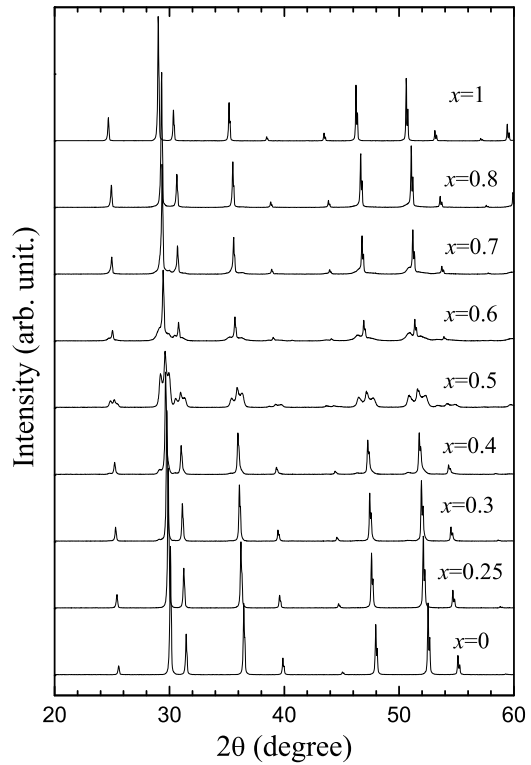}
\caption{\label{fig:xrd}XRD patterns for the
Cu$_{1-x}$Cd$_{x}$Ir$_{2}$S$_{4}$ samples quenched at 1373 K.}
\end{figure}

Fig.~\ref{fig:xrd} shows the XRD patterns for some of the
Cu$_{1-x}$Cd$_{x}$Ir$_{2}$S$_{4}$ samples quenched at 1373 K.
nearly single spinel phase is indicated for 0 $\leq x\lesssim $
0.4 and $0.7\lesssim x \leq1.0$. For $x$=0.5 and 0.6 samples,
however, there exist two additional broad peaks on both sides of
every diffraction position. Detailed analysis indicates that the
sample contains three spinel phases with different lattice
constants. The phase with the sharp XRD peaks is the metastable
solid-solution phase, obtained by the quenching. However, since
the quenching time was not short enough (Note that the sample was
sealed under vacuum in a quartz tube), the metastable
solid-solution phase decomposes into the Cd-rich and Cd-poor
phases during the cooling process. As the quenching temperature is
lowered, the Cd (or Cu) solubility decreases obviously. For
example, the Cd solubility is found to be 0.15 when quenched at
1123 K. So, the Cu/Cd substitution in the
Cu$_{1-x}$Cd$_{x}$Ir$_{2}$S$_{4}$ system exhibits miscibility gap
behavior.

\begin{figure}
\includegraphics[width=7cm]{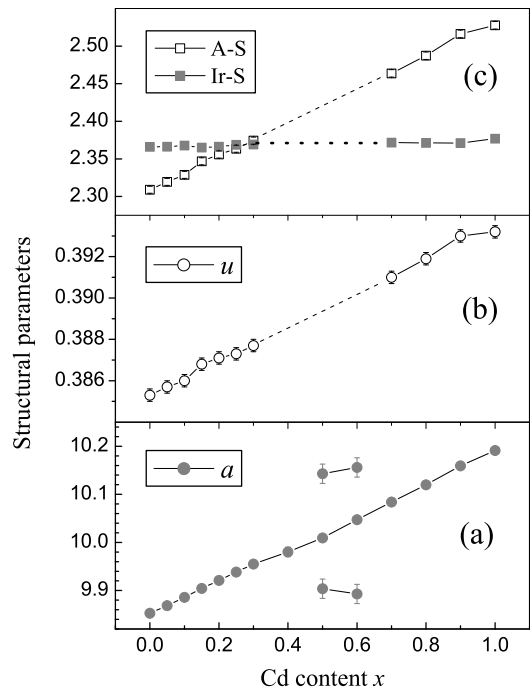}
\caption{\label{fig:structure}Lattice constant $a$ (a), parameter
$u$ (b), interatomic distances (c) as a function of Cd content $x$
in the Cu$_{1-x}$Cd$_{x}$Ir$_{2}$S$_{4}$ system. The unit of the
vertical axis is in \AA, except for the parameter $u$, which is
dimensionless.}
\end{figure}

The XRD data of the monophasic samples can be successfully fitted
by the Rietveld refinement method,~\cite{Izumi} which gives the
lattice constant $a$ and $u$ parameter for the thiospinel system.
It is also indicated that the Cu$_{1-x}$Cd$_{x}$Ir$_{2}$S$_{4}$
system crystalizes in \emph{normal} spinel structure, in which
Cu/Cd and Ir occupy A and B sites, respectively.
Fig.~\ref{fig:structure} shows the crystal structure parameters as
a function of Cd content. It can be seen that the lattice constant
increases linearly with increasing Cd substitution, obeying the
Vegard's law, except for the case of the additional phases
separated during the quenching process. The $u$ parameter, which
determines the atomic position of sulfur, also increases linearly.
As expected, the Cu/Cd-S (denoted as A-S) interatomic distance
increases remarkably due to the lattice expansion. However, the
Ir-S bond distance increase little, which is structurally related
to the stretching of the IrS$_{6}$ octahedra along [111]
directions. Compared with the result of the
Cu$_{1-x}$Zn$_{x}$Ir$_{2}$S$_{4}$ system,~\cite{Cao01} though the
Ir-S interatomic distance has no much difference against the
A-site substitution, the IrS$_6$ octahedra are obviously more
distorted in the present system.

\begin{figure}
\includegraphics[width=7cm]{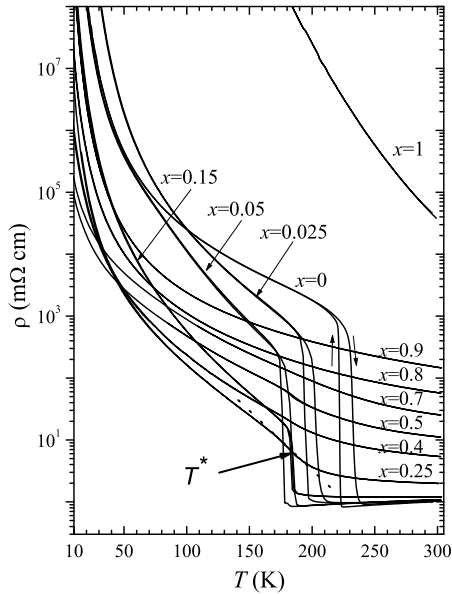}
\caption{\label{fig:resistivity}Temperature dependence of
resistivity for some Cu$_{1-x}$Cd$_{x}$Ir$_{2}$S$_{4}$ samples.
$T^{*}$ is defined as an inflexion temperature for the $\rho (T$)
curves.}
\end{figure}

Fig.~\ref{fig:resistivity} shows the temperature dependence of
resistivity in the Cu$_{1-x}$Cd$_{x}$Ir$_{2}$S$_{4}$ system. The
parent compound CuIr$_{2}$S$_{4}$ undergoes a MI transition at
$\sim$ 228 K with an obvious thermal hysteresis, consistent with
the earlier literatures.~\cite{Nagata94,Hagino,Furubayashi94} The
MI transition temperature $T_{MI}$ decreases to $\sim $ 185 K when
$x$=0.05. With further increasing the Cd substitution, the
$T_{MI}$ value does not decrease anymore. However, the thermal
hysteresis tends to disappear at this stage. At $x$=0.25, no
thermal hysteresis is detectable. The sudden change in resistivity
evolves into an inflexion for the $\rho (T$) curve at the
temperature $T^{*}$. This suggests a crossover from a first-order
transition to a higher-order one. When $x \ge $ 0.8, even an
inflexion cannot be observed in the whole temperature range. The
end member CdIr$_{2}$S$_{4}$ shows an insulating behavior. The
Arrhenius fitting for the data from 200 K to 300 K gives the
activation energy of 0.15 eV. So, the band gap is about 0.30 eV if
assuming that the sample is an intrinsic semiconductor. No
superconducting transition was observed down to 2 K.

It has been revealed that the low-temperature triclinic phase of
CuIr$_{2}$S$_{4}$ is a charge ordered insulator with Ir$^{4+}$
dimerizations.~\cite{Radaelli} Very recently, $d$-orbital
orientation ordering along Ir chains was suggested to account for
the x-ray absorption measurement result.~\cite{Croft} The model
shows the direct $d$-orbital overlap between Ir$^{4+}$, like the
case proposed in Ti$_{4}$O$_{7}$.~\cite{Lakkis,Cao01} If we
describe the Ir$^{4+}$-dimers as spin-singlet small bipolarons,
and regardless of the microscopic mechanism for the bipolaron
formation, the electronic state of the triclinic CuIr$_{2}$S$_{4}$
may be regarded as "bipolaron crystals".~\cite{Kobayashi} As
stated above, the thermal hysteresis associated with the
structural phase transition from cubic to triclinic disappears at
$x$=0.25. This rules out the possibility of the bipolaron crystals
for $x\geqslant$ 0.25 samples (Further low-temperature structural
analysis is needed to confirm this point). Noted that the
resistivity increases rapidly at the $T^{*}$ with decreasing
temperature. We suggest that this is due to the formation of
\emph{disordered} Ir$^{4+}$-dimers, which can be described as a
"bipolaron liquid".~\cite{Kobayashi} Theoretically, Emin
~\cite{Emin} studied the hopping transport for the singlet small
bipolarons. It was shown that one-electron transfers involving
small polarons coming from the bipolaron pair-breaking
\emph{always dominate} the dc conductivity. Therefore, when
bipolarons form at $T^{*}$, the concentration of the mobile
carriers decreases rapidly, resulting in the fast increase in the
resistivity.

Low-temperature conductivity behaviors may give information for
the insulating properties. Considering the formation of bipolarons
and three-dimensional variable-range-hopping (VRH) of localized
carriers,~\cite{Mott} we employed a two-channel conduction
equation,~\cite{Cao01}
\begin{equation}
\sigma=\sigma_{1}+\sigma_{2}=A \texttt{exp}(-E_{a}/k_{B}T)+B
\texttt{exp}[-(T_{0}/T)^{1/4}]. \label{eq1}
\end{equation}
The first term $\sigma_{1}$ describes the conductivity
contribution due to the thermally activated bipolaron
pair-breaking. The second term $\sigma_{2}$ expresses the VRH
conduction of the localized carriers. According to Emin's
description for strong small-bipolaron binding ,~\cite{Emin}
$E_{a}=(3\varepsilon_{b}+U)/4-|t|$, where $\varepsilon_{b}$, $U$
and $t$ denote the binding energy of small bipolarons, on-site
Coulomb repulsion and the transfer energy between sites,
respectively.

\begin{figure}
\includegraphics[width=8cm]{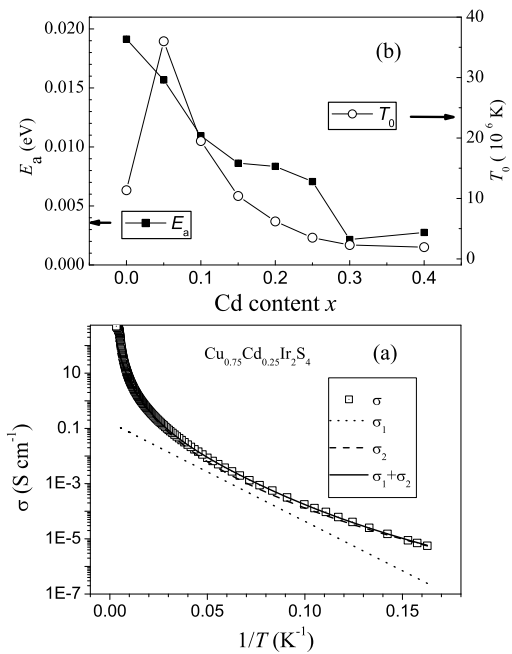}
\caption{\label{fig:conduct-fit}The fitting of low-temperature
conductivity: (a) Least-squared fitting using Eq. (1) for $x$=0.25
sample; (b) Fitted $E_{a}$ and $T_{0}$ values as a function of Cd
content. The lines are a guide to the eyes.}
\end{figure}

Fig. 4(a) shows the typical least-squared fitting for the $x$=0.25
sample using Eq. (1). It can be seen that the activated component
is needed to improve the fitting. Table~\ref{tab:table1} gives the
values of fitted parameters. To reveal the Cd-doping dependence,
$E_{a}$ and $T_{0}$ are plotted as a function of Cd content in
Fig. 4(b). One can see that $E_{a}$ decreases with increasing Cd
content. In the following analysis on the magnetic susceptibility,
$\varepsilon_{b}$ is shown to decease sharply at low Cd-doping,
then keeps almost unchanged. So, the rapid decrease in $E_{a}$ at
low Cd-doping is mainly due to the change of $\varepsilon_{b}$,
since $E_{a}=(3\varepsilon_{b}+U)/4-|t|$. Another parameter
$T_{0}$ in Fig. 4(b) is found to increase abruptly for small
Cd-doping, and then decrease monotonically as Cd content
increases. Similar result was also observed in the Zn-doped
system.~\cite{Cao01} The rapid increase of $T_{0}$ at low
Cd-doping may be due to enhanced localization resulting from
doping-induced disorder. As the doping level is increased,
however, the ratio of Ir$^{3+}$:Ir$^{4+}$=1:1 for a charge-ordered
state is not satisfied. One may expect that the density of
\emph{localized} states at the Fermi level increases with the Cd
doping, resulting in the decrease of $T_{0}$.~\cite{Mott}

\begin{table*}
\caption{\label{tab:table1}Parameters obtained by least-squared
fitting for the low-temperature (6 K $\sim$ 100 K) conductivity
data in Cu$_{1-x}$Cd$_{x}$Ir$_{2}$S$_{4}$ system using Eq. (1).
The numbers in the parentheses denote the uncertainty for the last
digital.}
\begin{ruledtabular}
\begin{tabular}{ccccc}
$x$&$A$ (S cm$^{-1}$)&$E_{a}$ (eV)&$B$ (10$^{6}$ S
cm$^{-1}$)&$T_{0}$ (10$^{6}$ K)\\ \hline
0&0.083(1)&0.0191(1)&3.1(1)&11.4(1)\\
0.05&0.071(3)&0.0157(1)&3200&36(2)\\
0.1&0.004(1)&0.0109(3)&600&19.5(4)\\
0.15&0.009(2)&0.0086(2)&95(6)&10.4(1)\\
0.2&0.07(1)&0.0084(2)&26(1)&6.2(1)\\
0.25&0.16(1)&0.0071(1)&5.0(1)&3.52(3)\\
0.3&0.013(1)&0.0022(1)&0.9(1)&2.3(1)\\
0.4&0.021(2)&0.0028(1)&0.51(4)&1.95(5)\\
\end{tabular}
\end{ruledtabular}
\end{table*}

Fig.~\ref{fig:susceptibility} shows the temperature dependence of
magnetic susceptibility for some Cu$_{1-x}$Cd$_{x}$Ir$_{2}$S$_{4
}$ samples. At the MI transition for CuIr$_{2}$S$_{4 }$, the
susceptibility has an abrupt change with a distinct thermal
hysteresis, because the Ir{$^{4+}$ ($S$=1/2) spin in the
low-temperature triclinic phase forms spin-singlet
dimers.~\cite{Radaelli} With increasing the Cd content, $T_{MI}$
first decreases ($x \le $ 0.05) rapidly and then keeps unchanged
(0.05 $ \le x \le $ 0.3), in confirmation with the above
resistivity result. Similarly, the thermal hysteresis disappears
at $x$=0.25. However, the susceptibility drop is still obvious at
$T^{*}$ $\sim$ 185 K, which supplies further evidence of the
bipolaron formation. When $x\geqslant0.8$, even the drop in $\chi$
is not detectable. For the insulating end member
CdIr$_{2}$S$_{4}$, the $\chi_{300K}$ value is $-6.4\times10^{-5}$
emu/mol, almost the same as the susceptibility of the
low-temperature insulating phase of
CuIr$_{2}$S$_{4}$.~\cite{Cao01}  No sign of superconductivity is
observed above 1.8 K for all the samples.

\begin{figure}
\includegraphics[width=7cm]{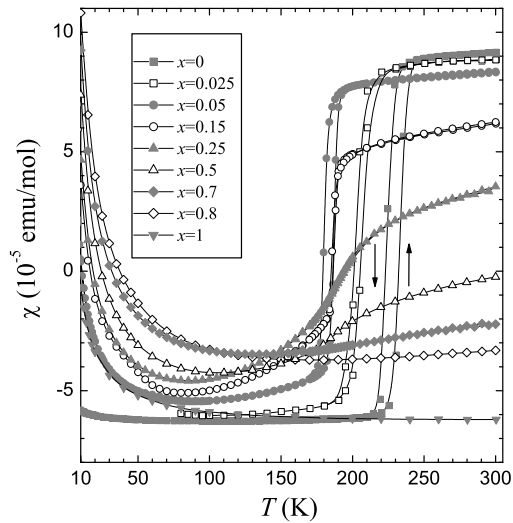}
\caption{\label{fig:susceptibility}Temperature dependence of
magnetic susceptibility for Cu$_{1-x}$Cd$_{x}$Ir$_{2}$S$_{4}$
samples. The arrows indicate the thermal hysteresis. The applied
field is 1000 Oe.}
\end{figure}

We note that the minimal susceptibility value for the samples
other than the end members is substantially larger than
$-6.4\times10^{-5}$ emu/mol. This cannot be explained only by the
impurity moments. Emin ~\cite{Emin} suggests that certain
paramagnetic susceptibility may be produced due to the thermal
activation of the small bipolarons. So, similar to our previous
treatment,~\cite{Cao01} the low-temperature susceptibility data
for the $x\leq$ 0.7 samples were analyzed using the equation below
\begin{equation}
\chi_{tot}=\chi_{0}+C_{1}/(T-\theta)+(C_{2}/T)
\texttt{exp}(-\Delta/k_{B}T).
\end{equation}
The first term on the right, including the core diamagnetism and
the Pauli paramagnetism $\chi_{Pauli}$, is assumed to be
temperature independent. The second term is the Curie-Weiss
paramagnetism part ($\chi_{CW}$), coming from impurities and/or
the lattice imperfections. The last term describes the
pair-breaking of the bipolarons ($\chi_{BP}$), where $\Delta $ is
the energy to activate \textit{one} localized spin. If the
exchange energy in a Ir$^{4+}$ dimer is $J$, we have,
$\varepsilon_{b}=2 \Delta=J-U$, where $U$ is the Coulomb repulsion
between the two carriers in a small bipolaron.

Fig. 6(a) shows the least-squared fitting for $x$=0.25 sample
using Eq. (2). It can be seen that the fitting is quite good. The
Curie-Weiss contribution mainly accounts for the susceptibility
upturn at low temperatures, while the $\chi_{BP}$ term elucidates
the temperature dependence of susceptibility just below $T^{*}$.
The fitted parameters are listed in Table~\ref{tab:table2}. As can
be seen, the $C_{1}$ value is rather small, corresponding to $\sim
0.1 \mu_{B}$. The $C_{2}$ data gives the effective moment $\mu_{J}
\sim 1.0 \mu_{B}$ for the pair-broken spin, and $\mu_{J}$ tends to
increase with increasing Cd content.

\begin{figure}
\includegraphics[width=7cm]{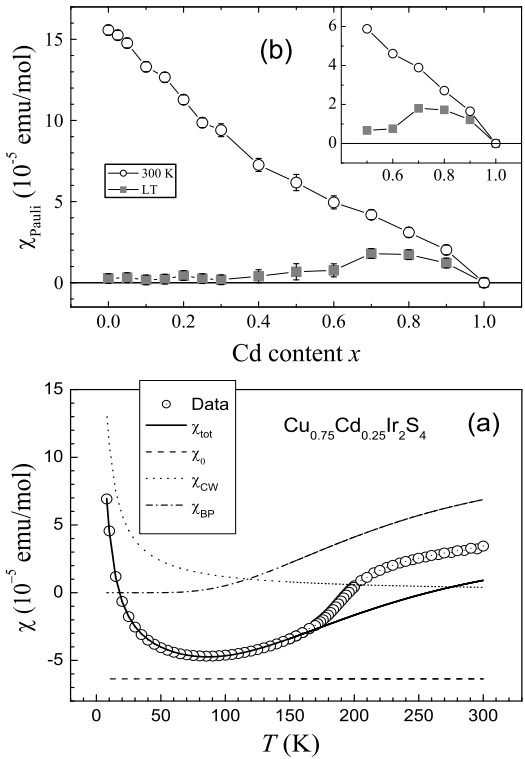}
\caption{\label{fig:chi-fit}(a) $\chi(T)$ data fitting for a
typical sample of $x$=0.25. (b) $\chi_{Pauli}$ at 300 K and at
room temperatures (LT) as a function of the Cd content. The inset
shows an enlarged plot.}
\end{figure}

\begin{table*}
\caption{\label{tab:table2}Parameters  from the fitting of the
low-temperature (10 K $\sim$ 160 K) susceptibility data in
Cu$_{1-x}$Cd$_{x}$Ir$_{2}$S$_{4}$ system using Eq. (2). The
numbers in the parentheses denote the maximum uncertainty for the
last digital.}
\begin{ruledtabular}
\begin{tabular}{cccccc}
$x$&$\chi_{0}$(10$^{-5}$ emu/mol)&$C_{1}$(emu K/mol)&$\theta$ (K)&$C_{2}$(emu K/mol)&$\Delta$ (K)\\
\hline
0&-6.35(1)&0.000071(1)&-3.6(1)&0.0395(7)&977(3)\\
0.025&-6.40(5)&0.00037(2)&-1(1)&0.069(1)&774(10)\\
0.05&-6.30(2)&0.00066(1)&-0.3(1)&0.061(1)&565(5)\\
0.1&-6.44(4)&0.00116(1)&-0.7(2)&0.095(1)&563(9)\\
0.15&-6.39(6)&0.00095(3)&-1.0(8)&0.130(1)&566(9)\\
0.2&-6.17(7)&0.00132(1)&-1.0(2)&0.126(5)&551(2)\\
0.25&-6.36(5)&0.00120(1)&-0.7(2)&0.130(2)&561(5)\\
0.3&-6.43(3)&0.00205(1)&-2.1(1)&0.118(1)&540(1)\\
\end{tabular}
\end{ruledtabular}
\end{table*}

By employing  the fitted parameters in Table~\ref{tab:table2}, the
$\chi_{Pauli}$ values in the high-temperature state can be roughly
extracted from the $\chi_{300K}$ values assuming that
$\chi_{300K}$=$\chi_{core}$+$\chi_{Pauli}$+$C_{1}/(300-\theta)$,
where the $\chi_{core}$ is approximately treated as the
$\chi_{300K}$ value of the insulator CdIr$_{2}$S$_{4}$. As shown
in Fig. 6(b), the room-temperature $\chi_{Pauli}$ value decreases
almost linearly, indicating that the density of state at $E_{F}$
decreases with increasing Cd content since
$\chi_{Pauli}=\mu_{B}^{2}N(E_{F})$. Therefore, one may conclude
that the Cd substitution fills the hole carriers, resulting in the
increase of resistivity and the decrease of magnetic
susceptibility. Compared with the result of the Zn
substitution,~\cite{Cao01} the present Cd substitution leads to
sharper decrease in the $N(E_{F})$, implying that valence band
width of the Cd-substituted system is narrower. This is consistent
with the lattice expansion due to the Cd substitution.

The $\chi_{Pauli}$ values in the low-temperature state, if exists,
can also be estimated in a similar way. It is shown in Fig. 6(b)
that the low-temperature $\chi_{Pauli}$ value keeps nearly zero
for $x\lesssim0.3$, consistent with the picture of dimerizations
of Ir$^{4+}$. This result is in sharp contrast with that of
Cu$_{1-x}$Zn$_{x}$Ir$_{2}$S$_{4}$ system. The latter system shows
that $\chi_{0}$ changes rapidly from -6.2$\times10^{-5}$ emu/mol
for $x$=0.1 to -1.5$\times10^{-5}$ emu/mol for $x$=0.25, which was
explained by the fact of two-phase coexistence.~\cite{Cao01}
Therefore, it is very likely that the present Cd-doping system
keeps only one phase at low temperatures for $x\lesssim0.3$.
Low-temperature XRD measurement is needed for confirming this
point. When $x\geq 0.7$, however, the low-temperature
$\chi_{Pauli}$ is remarkably increased, indicating that the
bipolarons become unstable. This is consistent with the fact that
the $x\geqslant0.8$ samples shows no transition into the bipolaron
state at all in the whole temperature range. Therefore, the solid
solution in the range of 0.8 $<x<$ 1 can be regarded as a
polaronic semiconductor. In the case of 0.4 $ \le x \le $ 0.6, the
non-zero $\chi_{Pauli}$ is observed, which is probably originated
from the contribution of Cd-rich phase formed during the quenching
process.

Fig.~\ref{fig:delta} shows the fitted $\Delta$ values as well as
the transition temperature $T_{t}$ ($T_{MI}$ or $T^{*}$) for
Cu$_{1-x}$Cd$_{x}$Ir$_{2}$S$_{4}$ system. It can be seen that
$\Delta$ changes in the same tendency as the $T_{t}$ does. This
indicates that the formation of the Ir$^{4+}$-dimers or the
bipolarons is closely relevant to the first-order or the
second-order transitions at $T_{t}$.

\begin{figure}
\includegraphics[width=7cm]{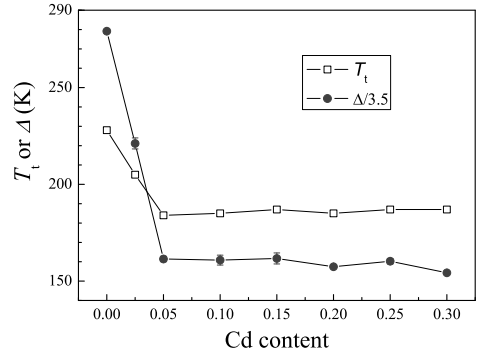}
\caption{\label{fig:delta}$T_{t}$ and $\Delta$ as a function Cd
content in Cu$_{1-x}$Cd$_{x}$Ir$_{2}$S$_{4}$ system.}
\end{figure}

The result of heat capacity measurement is shown in
Fig.~\ref{fig:capacity}. For $x$=0.15, a sharp peak can be seen at
the $T_{MI}$, accompanied with a thermal hysteresis. This is
consistent with the first-order phase transition. The enthalpy and
entropy of the transition are 450 J mol$^{-1}$ and 2.43 J K$^{-1}$
mol$^{-1}$, respectively. When $x$=0.25, a broad peak can be seen
at the $T^{*}$ on the background of the lattice specific heat, but
no thermal hysteresis was observed. This observation indicates
that the transition is of second order. In the case of $x$=0.5,
detectably small peak appears at $\sim$ 185 K. For $x$=0.8, no
second order transition can be found. By using the Debye theory,
the Debye temperature $\Theta_{D}$ was determined to be about 100
K, independent of the Cd content.

\begin{figure}
\includegraphics[width=7cm]{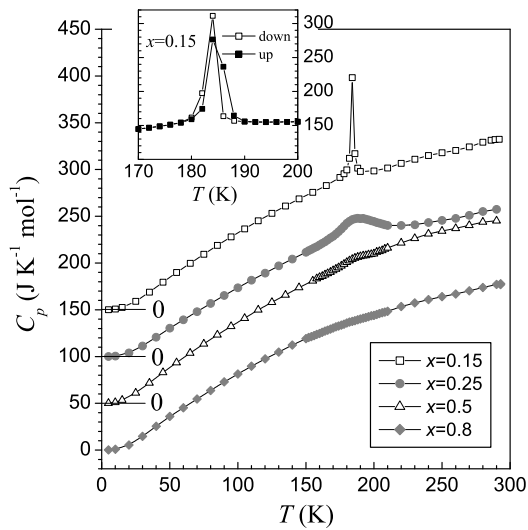}
\caption{\label{fig:capacity}Temperature dependence of the
specific heat in the Cu$_{1-x}$Cd$_{x}$Ir$_{2}$S$_{4}$ system. In
order to distinguish each curve, the data were shifted one by one
along the vertical axis.}
\end{figure}

Based upon the above result, the phase diagram of
Cu$_{1-x}$Cd$_{x}$Ir$_{2}$S$_{4}$ can be concluded as shown in
Fig. 9(a). In the low Cd-doping region, the high temperature phase
is a cubic metal. The low temperature phase is a triclinic
insulator with charge ordering and Ir$^{4+}$-dimerization, called
bipolaron crystals. It is noted that the MI transition temperature
first drops rapidly, then remains unchanged. For the intermediate
Cd-doping ($0.2<x<0.8$), the first-order transition is changed
into a second order one at the $T^{*}$. It is noted that the
$T^{*}$ is almost independent of Cd content. This may be relevant
to the fact that Ir-S bond distance hardly changes with the Cd
content. The phase above the $T^{*}$ becomes a kind of
semiconductor as called a polaronic semiconductor, while the phase
below the $T^{*}$ can be described as a bipolaron liquid. It is
noted that the present samples of $x$=0.5 and 0.6 contain multiple
spinel phases due to the miscibility-gap effect. In the case of
heavy Cd substitution ($0.8<x<1$), the hole concentration is so
low that bipolaron is not stable anymore. Furthermore, with
increasing the hole filling, the Fermi level may move across the
mobility edge,~\cite{Mott} locating in the Anderson localization
state. So, the sample shows a semiconducting behavior. The end
member CdIr$_{2}$S$_{4}$ is a band insulator.

\begin{figure}[tbp]
\includegraphics[width=7cm]{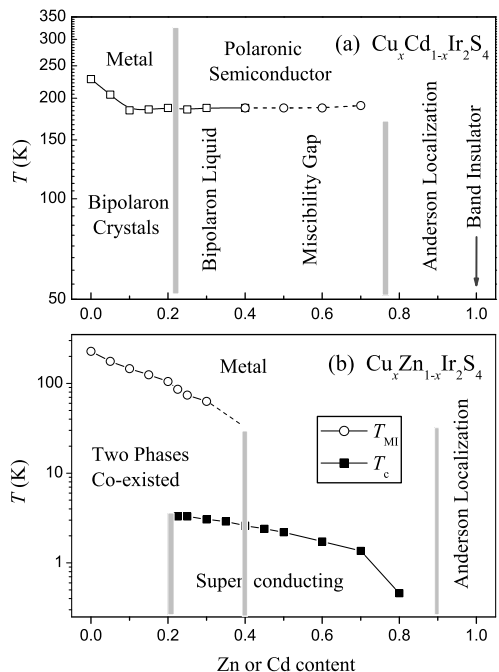}
\caption{\label{fig:diagram}(a) Electronic phase diagram in the
Cu$_{1-x}$Cd$_{x}$Ir$_{2}$S$_{4}$ pseudo-binary spinel system. (b)
The phase diagram of Cu$_{1-x}$Zn$_{x}$Ir$_{2}$S$_{4}$ is also
shown for comparison.~\cite{Cao01}}
\end{figure}

The phase diagram differs much from that of an analogue
Cu$_{1-x}$Zn$_{x}$Ir$_{2}$S$_{4}$ system shown in Fig.
9(b).~\cite{Cao01} One striking difference is that the Zn-doping
induces superconductivity below 3 K, while the Cd-doping does not
lead to any superconducting ground state. Instead of the
superconducting Cooper pairs, small bipolarons become stable in
the range of 0.25 $ \le x \le $ 0.8 for the Cd-substituted system.
Theoretically, Chakraverty~\cite{Chakraverty} earlier showed that
a Cooper-pair superconductor may change into a bipolaronic
insulator when the electron-phonon coupling constant goes beyond a
critical value. It was also stressed that the coupling constant
increases with decreasing the bandwidth in the strong coupling
regime. Therefore, the absence of superconductivity in the
Cd-substituted system is possibly ascribed to the relatively small
bandwidth arising from the negative chemical pressure by the Cd
substitution. Nevertheless, the physical pressure effect in
Cu$_{1-x}$Zn$_{x}$Ir$_{2}$S$_{4}$ system~\cite{Cao03} reveals the
opposite tendency. The applied pressure, which is generally
assumed to increase the bandwidth, induces a
superconductor-to-semiconductor transition. The low-temperature
semiconducting phase was also associated with the bipolaron
formation. Such contradiction deserves further investigations.

Another difference concerns the electrical conduction for the
intermediate doping. The semiconducting behavior above $T^{*}$ in
Cu$_{1-x}$Cd$_{x}$Ir$_{2}$S$_{4}$ ($0.2<x<0.8$) seems to be rather
unusual. We note that the $\chi_{Pauli}$ value of the present
system is much smaller than the corresponding one in
Cu$_{1-x}$Zn$_{x}$Ir$_{2}$S$_{4}$. This means that the density of
state at $E_{F}$ is much lower in the Cd-doped system. However, to
elucidate the semiconducting behavior, one may need to consider
the distortion of the IrS$_{6}$ octahedra as well as the Anderson
localization effect.

\section{\label{sec:level4}CONCLUSION}

In summary, the Cd substitution for Cu in CuIr$_{2}$S$_{4}$ system
was studied for understanding the variation of the MI transition
and the effect of chemical pressures on the charge ordering and
the Ir$^{4+}$ dimerizations. The XRD structural analysis indicates
that the Cd substitution increases the lattice constant and the
distortion of IrS$_{6}$ octahedra obviously, but affects the Ir-S
bond distance quite little. By the measurement of temperature
dependence of electrical resistivity, magnetic susceptibility and
specific heat, the electronic phase diagram has been tentatively
established, which is quite different from an analogue system of
Cu$_{1-x}$Zn$_{x}$Ir$_{2}$S$_{4}$. It is found that the
first-order MI transition at 228 K in the parent compound
CuIr$_{2}$S$_{4}$ is at first suppressed to a level of 185 K, and
then changed into a second-order transition for $x>$ 0.2. The
second-order transition disappears when $x\ge$ 0.8. The end-member
compound CdIr$_{2}$S$_{4}$ is shown as an insulator with a
band-gap of 0.3 eV. Both the electrical resistivity and the
magnetic susceptibility data suggest that the phase below the
$T^{*}$ can be described as a bipolaron liquid. The absence of
superconductivity in the present system can thus be understood in
terms of the transition from the BCS Cooper pairs to small
bipolarons due to the decrease of the bandwidth resulted from the
negative chemical pressures.

\begin{acknowledgments}
   We are indebted to Dr. T. Sato for the chance to use the SQUID
magnetometer. One of the authors (Cao) acknowledges the support
from NSFC with the Grant No. 10104012.
\end{acknowledgments}

\end{document}